\documentclass{jpsj-suppl}
\usepackage{txfonts} 

\title{Assembly and Bench Testing of a Spiral Fiber Tracker for the J-PARC TREK/E36 Experiment}

\author{Makoto \textsc{Tabata}$^{1,*}$, S\'{e}bastien \textsc{Bianchin}$^{2}$, Michael D. \textsc{Hasinoff}$^{3}$, Robert S. \textsc{Henderson}$^{2}$, Keito \textsc{Horie}$^{4}$, Youichi \textsc{Igarashi}$^{5}$, Jun \textsc{Imazato}$^{5}$, Hiroshi \textsc{Ito}$^{1}$, Alexander \textsc{Ivashkin}$^{6}$, Hideyuki \textsc{Kawai}$^{1}$, Yury \textsc{Kudenko}$^{6}$, Oleg \textsc{Mineev}$^{6}$, Suguru \textsc{Shimizu}$^{4}$, Akihisa \textsc{Toyoda}$^{5}$, and Hirohito \textsc{Yamazaki}$^{7}$}

\inst{$^{1}$Department of Physics, Chiba University, Chiba, Japan\\
$^{2}$Canada's National Laboratory for Particle and Nuclear Physics (TRIUMF), Vancouver, Canada\\
$^{3}$Department of Physics and Astronomy, University of British Columbia, Vancouver, Canada\\
$^{4}$Department of Physics, Osaka University, Toyonaka, Japan\\
$^{5}$Institute of Particle and Nuclear Studies (IPNS), High Energy Accelerator Research Organization (KEK), Tsukuba, Japan\\
$^{6}$Institute for Nuclear Research (INR) of the Russian Academy of Sciences (RAS), Moscow, Russia\\
$^{7}$Research Center for Electron Photon Science, Tohoku University, Sendai, Japan}

\email{makoto@hepburn.s.chiba-u.ac.jp}

\abst{This study presents the recent progress made in developing a spiral fiber tracker (SFT) for use in the experiment TREK/E36 planned at the Japan Proton Accelerator Research Complex. This kaon decay experiment uses a stopped positive kaon beam to search for physics beyond the Standard Model through precision measurements of lepton universality and through searches for a heavy sterile neutrino and a dark photon. Detecting and tracking positrons and positive muons from kaon decays are of importance in achieving high-precision measurements; therefore, we designed and are developing the new tracking detector using a scintillating fiber. The SFT was completely assembled, and in a bench test, no dead channel was determined.}

\kword{tracker, scintillating fiber, kaon decay, lepton universality, J-PARC TREK/E36}

\begin{document}
\maketitle

\section{Introduction}

We are currently developing a charged particle tracking detector, known as the spiral fiber tracker (SFT) \cite{cite1}, for use in the E36 experiment \cite{cite2,cite3,cite4} scheduled at the K1.1BR beam line in the Hadron Experimental Facility of the Japan Proton Accelerator Research Complex (J-PARC). This positive kaon decay experiment will search for physics beyond the Standard Model by testing lepton flavor universality and searching for a heavy sterile neutrino and a dark photon \cite{cite5,cite6,cite7,cite8}. To search for a violation of lepton universality, we focus, in particular, on precisely measuring the ratio of the kaon decay widths $R_{\rm K} = \Gamma (K^+ \to ~e^+\nu)/\Gamma (K^+ \to ~\mu ^+\nu )$ using a stopped kaon beam.

\begin{figure}[t] 
\centering 
\includegraphics[width=0.837\textwidth,keepaspectratio]{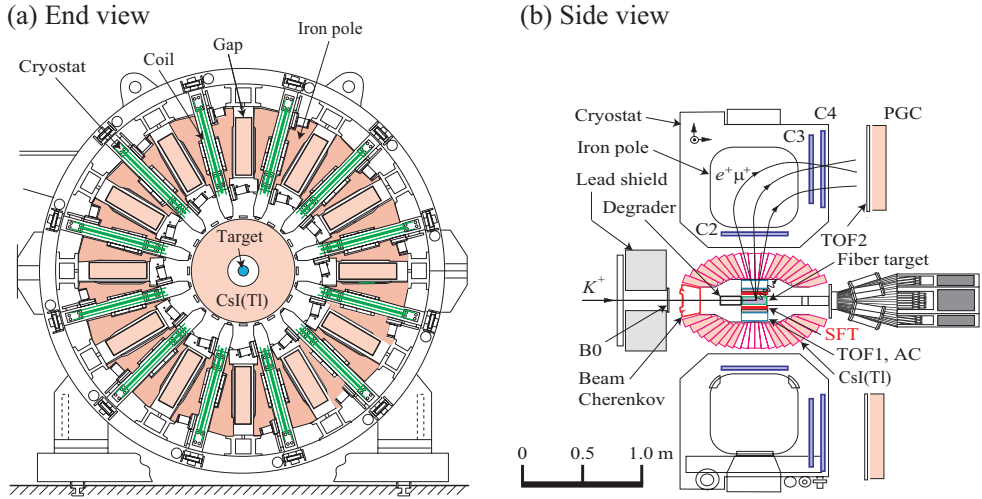}
\caption{(a) End- and (b) side-views of the TREK/E36 detector system.}
\label{fig:fig1}
\end{figure}

For this experiment, we are building a new TREK/E36 detector system (Fig. \ref{fig:fig1}) by upgrading the experiment E246 apparatus \cite{cite9,cite10}, which was based on a twelve-sector superconducting toroidal spectrometer that was previously used at the High Energy Accelerator Research Organization (KEK), Tsukuba, Japan. Conducting high-precision measurements depends on efficiently identifying and tracking charged particles (i.e., positrons and positive muons) from kaon decays. Particle identification is performed by measuring the time-of-flight (TOF) between the TOF1 and TOF2 scintillation counters, threshold aerogel Cherenkov (AC) counters with a refractive index of 1.08, and lead (Pb) glass Cherenkov (PGC) counters for robust analysis. As shown in Fig. \ref{fig:fig1}(b), the TOF1 and AC counters surround the kaon stopping active target made of plastic scintillating fibers \cite{cite11}, and the TOF2 and PGC counters are located at the exit of the spectrometer.

To reliably determine the momentum of the charged particles requires at least four-point tracking, which suggests that segments must be tracked before and after the magnetic field generated by the spectrometer. Four independent tracking devices allow us to measure the detection efficiency of the particles by three-of-four-point tracking. The E246 detector has three layers of multiwire proportional chambers (MWPCs). One layer (i.e., C2) and the other two layers (i.e., C3 and C4) of MWPCs are arranged upstream and downstream of the magnetic field, respectively (Fig. \ref{fig:fig1}(b)). Consequently, an additional tracking device (i.e., C1) is required upstream of the magnetic field. Therefore, we decided to develop and install the SFT in the very limited space around the target at the center of the TREK/E36 detector. We expect its detection efficiency to be higher than 98\%.

\section{Design of Spiral Fiber Tracker}

\begin{figure}[t] 
\centering 
\includegraphics[width=0.451\textwidth,keepaspectratio]{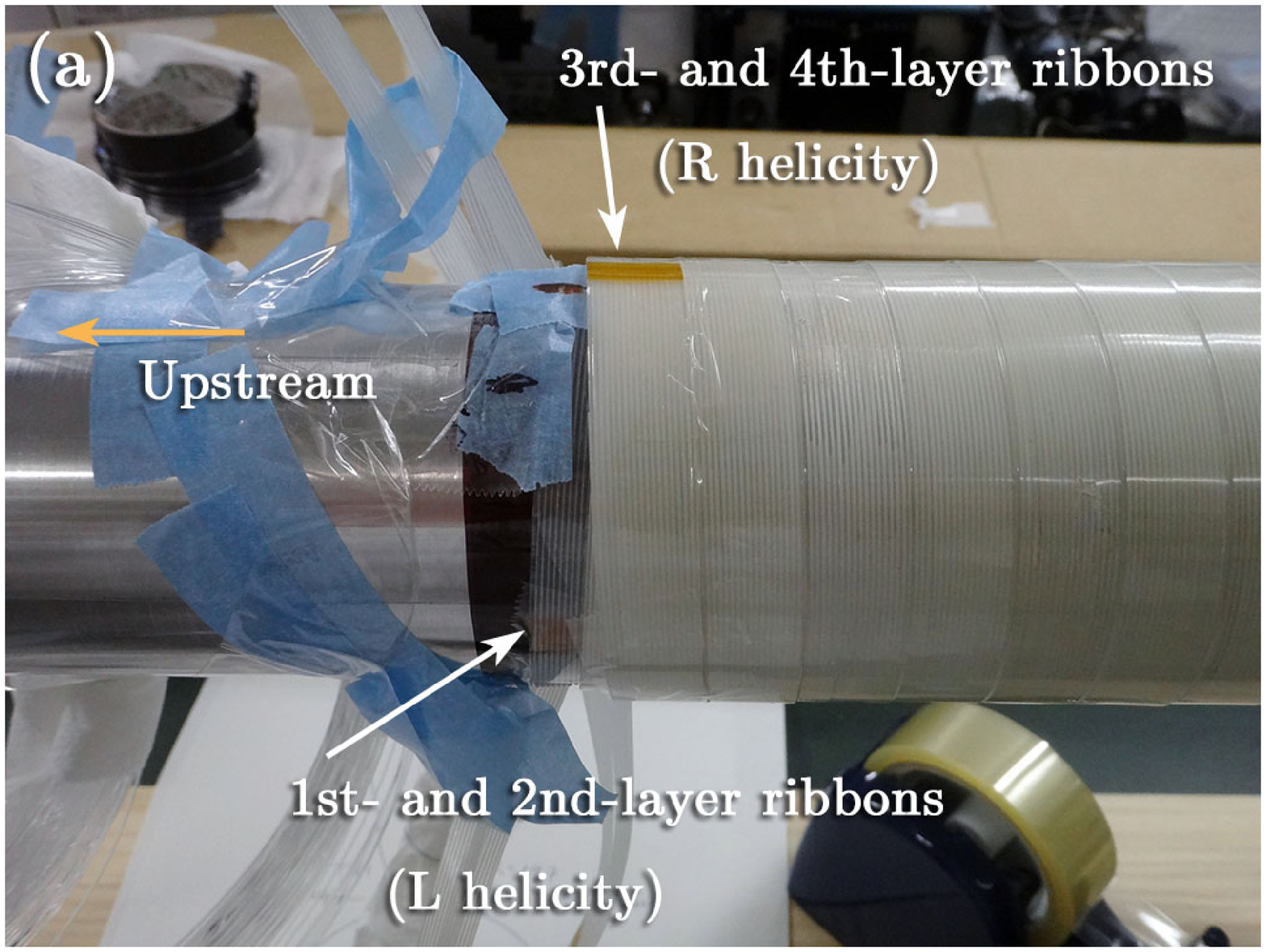}
\includegraphics[width=0.451\textwidth,keepaspectratio]{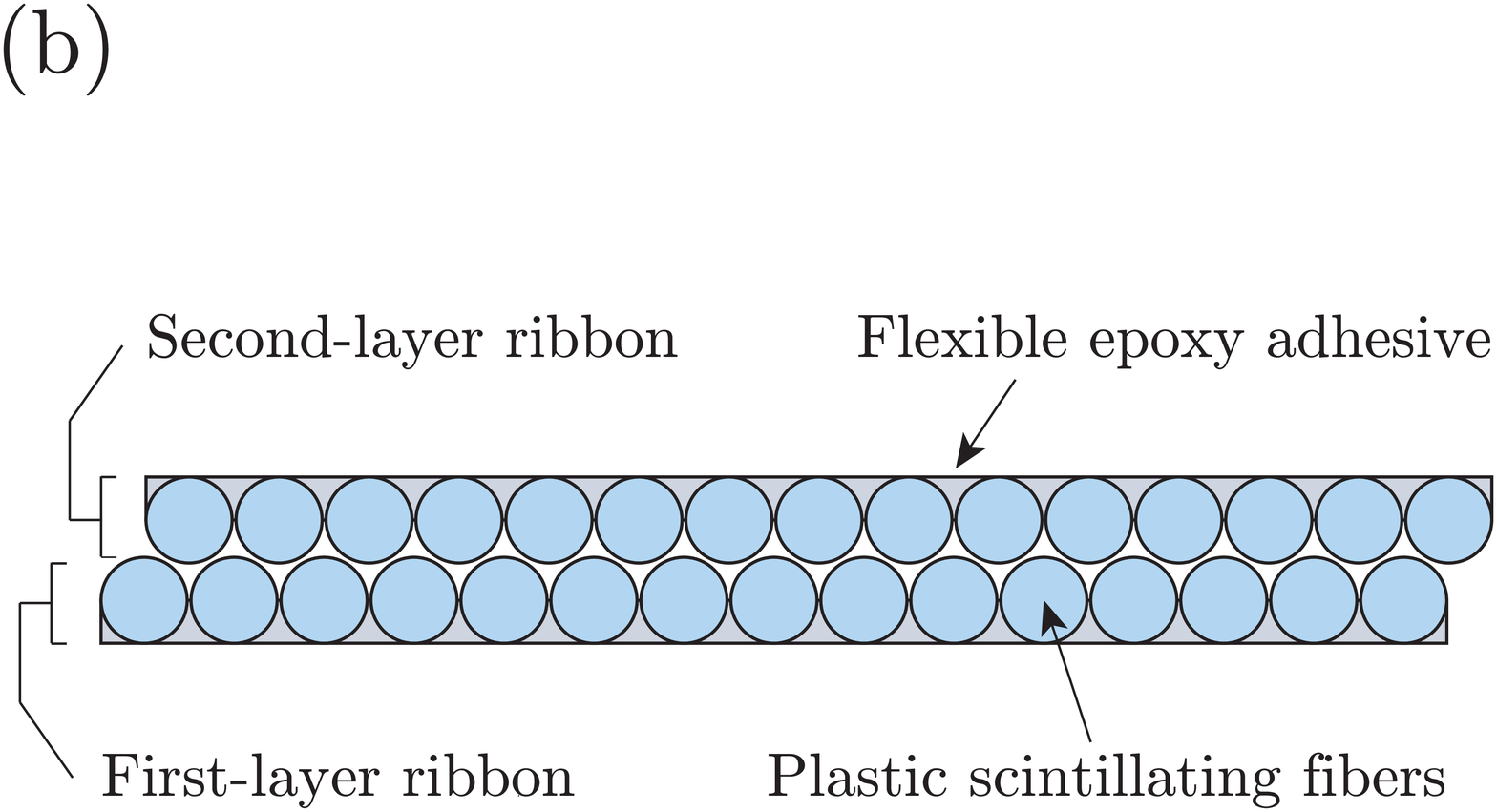}
\caption{(a) Close-up of spiral fiber tracker. The inner (outer) fiber ribbon is coiled with left (right) helicity viewed from downstream. (b) Cross-sectional view of staggered fiber configuration. One side (outside of staggered combination) of each fiber ribbon is glued using an epoxy adhesive.}
\label{fig:fig2}
\end{figure}

The heart of the SFT consists of multilayer fiber ribbons, each containing a single layer of 1-mm-diameter standard-core-type multiclad plastic scintillating round fibers (SCSF-78MJ, Kuraray Co., Ltd., Japan) \cite{cite12}. The peak wavelength of scintillating light emitted from the fiber is 450 nm. The conceptual design of the SFT can be seen in Fig. \ref{fig:fig2}(a). A total of four layers of ribbons, forming two sets of two layers in a staggered fiber configuration (Fig. \ref{fig:fig2}(b)), are spirally wound around the 79-mm-diameter kaon stopping target holder (approximately 20 cm long) integrated with a TOF1/AC support structure made of aluminum. The bending loss of light transmission of the fiber is below 5\% for a bending diameter of 80 mm \cite{cite12}.

Fifteen (inner) or seventeen (outer) fibers are glued to form a single-layer flat ribbon approximately 5 m long. The attenuation length of the fiber at 254 nm wavelength ($\lambda $) is longer than 4 m \cite{cite12}. The two-layer combinations are coiled in two helices with an intersection angle and have different ribbon widths (approximately 15 and 17 mm) to phase shift the candidate positions of charged particle hits; those candidate positions are where the inner and outer fibers that detect scintillation photons cross each other in each ribbon turn. Specifically, depending on ribbon turns, the candidate hit positions have different azimuthal angles around the beam axis.

At both ends of each fiber, scintillation photons created in the fibers are read by 128 multipixel photon counters (MPPCs; S10362-11-050C, Hamamatsu Photonics K.K., Japan). For each fiber, approximately 4-m-long (2-m-long) clear optical fibers are employed to extend, with low transmission losses, the scintillating fibers at the upstream (downstream) side to the MPPCs. The fiber--MPPC connection light-tight module is shared with the kaon stopping fiber target and is cooled. To determine the actual hit position from the candidate positions (i.e., ribbon turns), we shall use supplementary information from tracking on the active fiber target, in addition to timing information from both ends of the SFT readout.

\section{Basic Performance Measurements of Prototype Fiber Ribbon}

In March 2014, we bench tested a prototype scintillating fiber ribbon glued by the Moderation-Line Co., Ltd., Japan, for basic performance by measuring its charged particle detection efficiency. At that time, only a single test-produced ribbon was available. It was approximately 1.5 m long and consisted of sixteen 1-mm-diameter plastic scintillating fibers. A span of approximately 1.2 m in the middle of the fibers was glued using polyurethane adhesive. At both ends of the ribbon, where the fibers were not glued, the fiber ends were bundled by school glue, well polished with sandpaper, and connected to two photomultiplier tubes (PMTs; R9880U-210, Hamamatsu Photonics K.K.). The typical quantum efficiency of the PMTs is approximately 33\% at $\lambda $ = 450 nm (catalog value). The MPPC readout was not ready for this test.

\begin{figure}[t] 
\centering 
\includegraphics[width=0.451\textwidth,keepaspectratio]{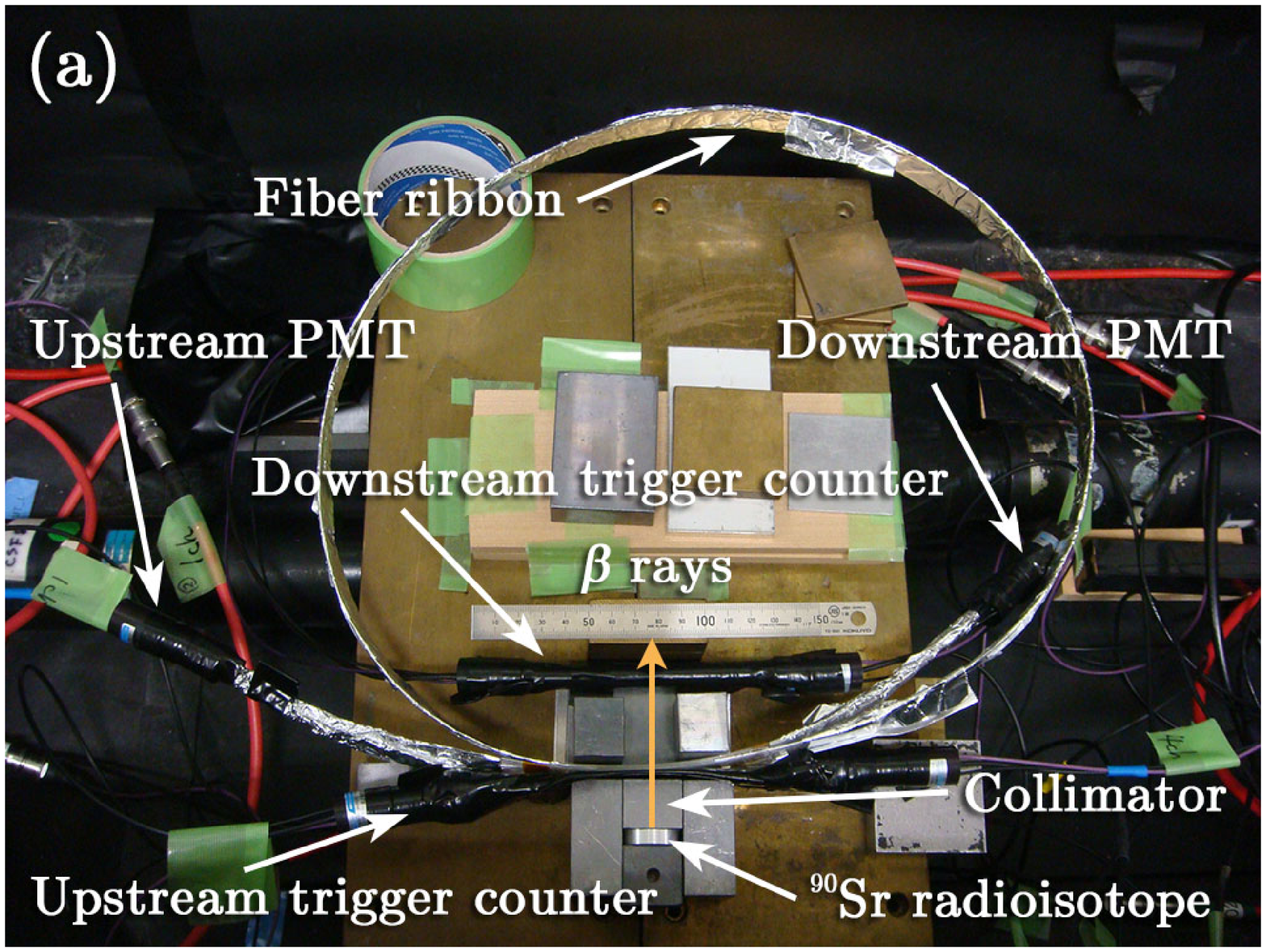}
\includegraphics[width=0.451\textwidth,keepaspectratio]{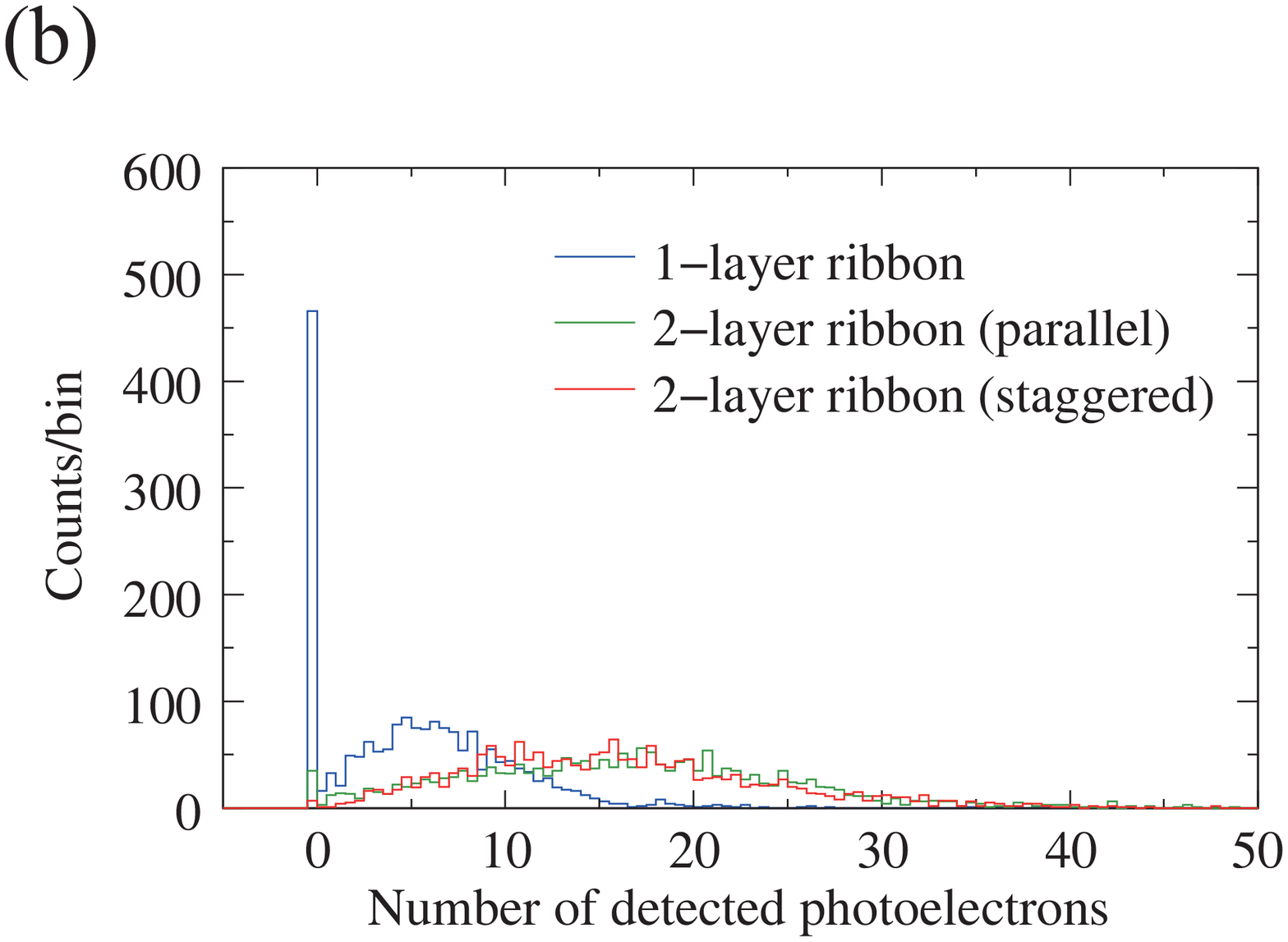}
\caption{(a) Experimental setup in a light-shielded box for the two-layer fiber ribbon. (b) Number distribution of photoelectrons detected by downstream PMT for (blue) single-layer ribbon, (green) two-layer ribbon in parallel fiber configuration, and (red) two-layer ribbon in staggered fiber configuration. For each fiber configuration, 1900 events were collected. The bin width was 0.5 photoelectrons.}
\label{fig:fig3}
\end{figure}

As a source of minimum ionizing particles, we used $\beta $ rays generated from a strontium-90 radioisotope. The experimental setup inside a light-shielded box is shown in Fig. \ref{fig:fig3}(a). The fiber ribbon was exposed to $\beta $ rays through a 5-mm-hole collimator fabricated from a 2-cm-long lead block. Event triggers were generated by coincidence signals from two scintillation counters positioned upstream and downstream of the ribbon. The upstream and downstream scintillation counters consisted of a multilayer sheet comprising 0.2-mm-diameter fibers and a 5-mm-thick block, respectively. These served as plastic scintillators.

Analog-to-digital converter spectra were obtained by CAMAC for single- and two-layer ribbons. First, to test the single-layer fiber ribbon, we irradiated it with $\beta $ rays from a source positioned approximately 20 cm from one fiber terminal. Here, the PMT connected to the near end of the fiber and the PMT connected to the opposite end of the fiber are called the ``upstream'' and ``downstream'' PMTs, respectively (Fig. \ref{fig:fig3}(a)). Next, the two-layer part of the ribbon was produced by winding it so that the ribbon formed a cylinder. In the two-layer part, the ribbon was basically overlapped in a staggered fiber configuration. As a reference, it was also overlapped in a parallel fiber configuration and measured.

On average, 15.5 photoelectrons were detected in the two-layer ribbon of the staggered fiber configuration. Fig. \ref{fig:fig3}(b) shows the number distribution of photoelectrons detected by the downstream PMT. For a threshold of three photoelectrons at the upstream or downstream PMT, we obtained a detection efficiency of 99.5\% for the two-layer staggered fiber configuration. High detection efficiency was basically confirmed in this fiber configuration. In contrast, the detection efficiencies were 97.5\% and 74.1\% for the two-layer parallel fiber configuration and single-layer ribbon, respectively. These low efficiencies can be attributed to events that occurred when $\beta $ rays passed through the dead zones of fiber claddings.

\section{Fiber Assembly}

In April 2014, we assembled the SFT at KEK. In advance of assembly, fifteen or seventeen plastic scintillating fibers were glued using flexible epoxy adhesive (unlike the prototype) to form a total of four ribbons. A welding technique was used by the Moderation-Line Co., Ltd., to connect all fiber ends to clear optical fibers. Light transmission losses at the two connections between the scintillating and clear fibers (i.e., the connection points from the clear fibers to the scintillating fibers and the reverse) were measured by the company using a red laser diode ($\lambda $ = 635 nm). This allows light leaks to be detected visually. For all 64 fibers, the average and maximum transmission losses of light through the two connection points were 23\% and 29\%, respectively, while the nominal value was 28\%.

\begin{figure}[t] 
\centering 
\includegraphics[width=0.451\textwidth,keepaspectratio]{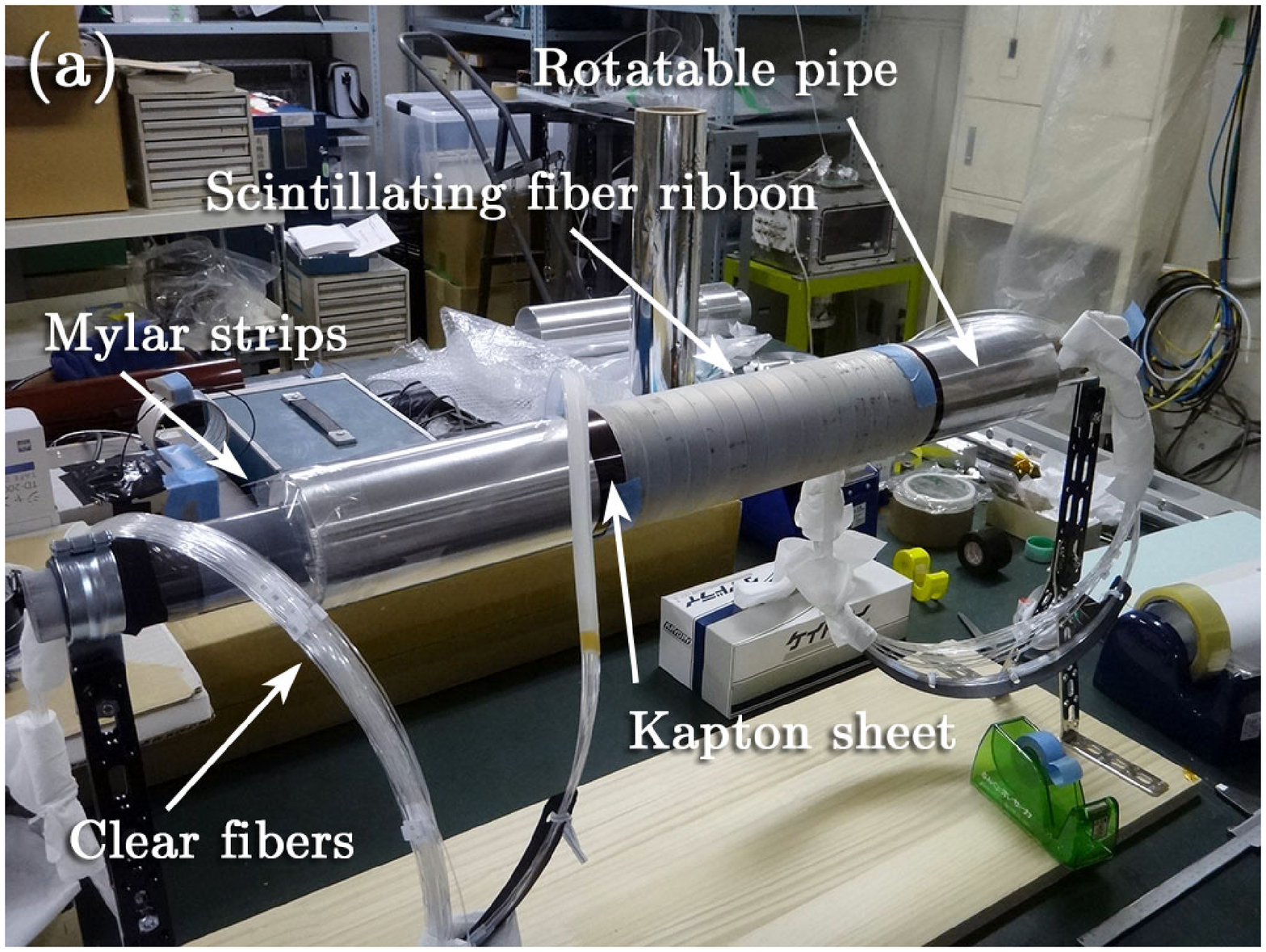}
\includegraphics[width=0.451\textwidth,keepaspectratio]{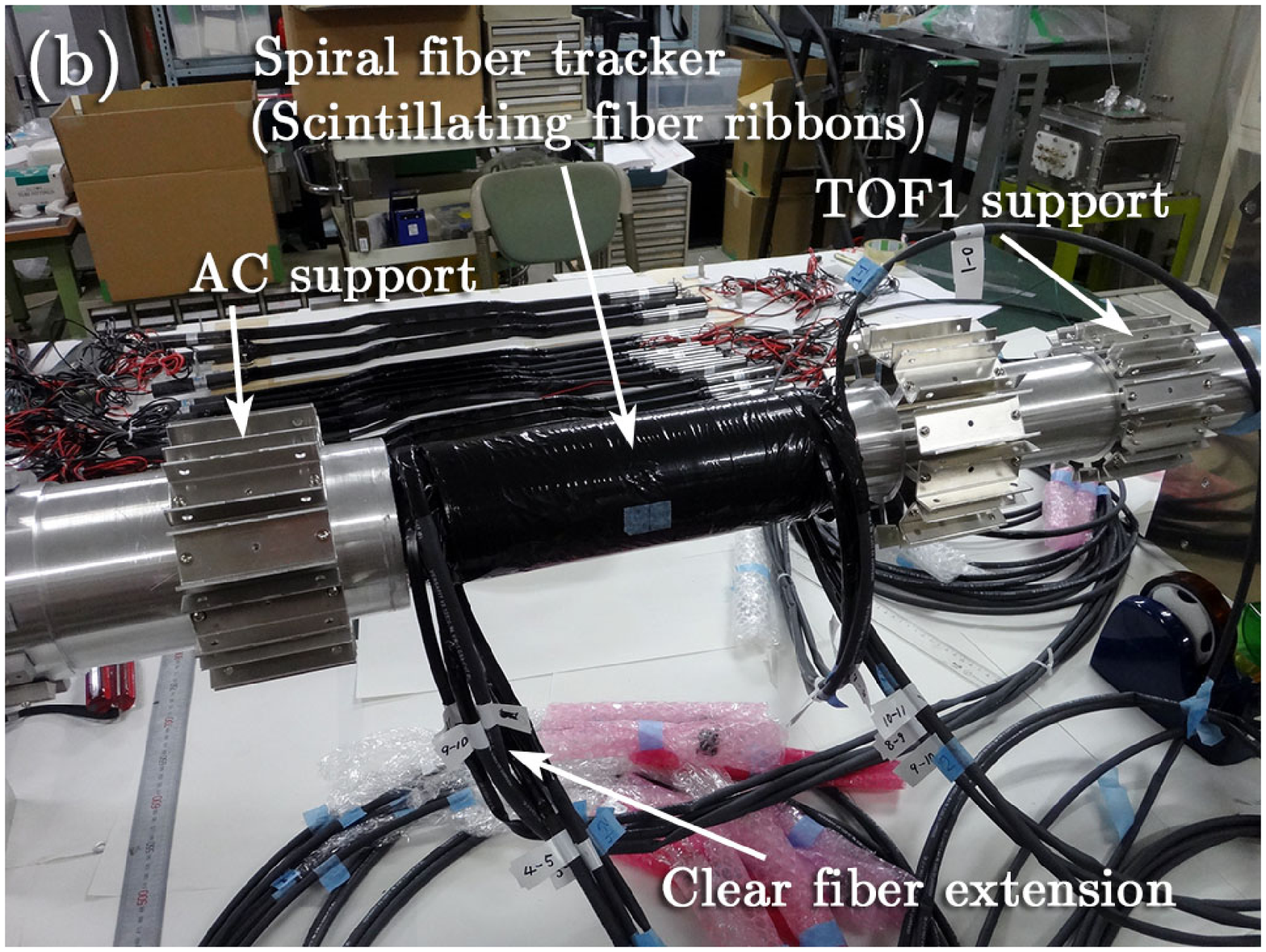}
\caption{(a) Spiral fiber tracker just after the first-layer ribbon was coiled around the dummy pipe. (b) Assembled spiral fiber tracker around the kaon stopping target holder.}
\label{fig:fig4}
\end{figure}

For the ease of assembly on a desktop, we prepared a 60-cm-long rotatable aluminum pipe with the same diameter (i.e., 79 mm) as the actual kaon stopping target holder (Fig. \ref{fig:fig4}(a)). To coil fiber ribbons on it, the dummy pipe was doubly wrapped with a 30-cm-long Kapton sheet having 150 $\mu $m thickness. To facilitate the removal of the coiled ribbons from the pipe, eight Mylar strips with 100 $\mu $m thickness were placed between the pipe and the Kapton sheet along the longitudinal direction of the pipe. Next, the first-layer fiber ribbon was coiled around the Kapton sheet with left helicity as viewed from downstream. The coiled ribbon was fixed by Mylar tape. The second-layer ribbon was then coiled on the first-layer ribbon in the staggered fiber configuration. To avoid the overlap of fiber extraction through the TOF1 and AC counters, the starting point of the second-layer ribbon was set to a different azimuthal angle around the beam axis relative to the starting point of the first-layer ribbon. Similarly, the third- and fourth-layer ribbons were coiled with right helicity over the previous two layers. Finally, the coiled ribbons and extended clear fibers were shielded from light with a black resin sheet and heat-shrinkable black tubes (nonheated) with diameters of 7 mm.

To connect the fibers to the MPPCs, each clear fiber edge was glued to a (positive) coupler using optical cement (EJ-500, Eljen Technology, USA). The MPPCs were equipped with negative couplers. Using 16 dedicated jigs, a total of 32 fiber ends could be glued to the couplers at one time. The gluing was aged for one day. Redundant optical cement around the fiber ends was removed by polishing the fiber termini. After the fiber termini with the couplers were roughly polished with P2000 sandpaper, they were polished using a metal polish and then wiped with ethanol.

The coiled ribbon body with the Kapton sheet was then transferred from the dummy pipe to the actual target holder. After removing the Mylar strips between the pipe and the Kapton sheet, the ribbon body was easily pulled out, leaving the pipe. Finally, the ribbon body was installed in the prepared location around the target holder, as shown in Fig. \ref{fig:fig4}(b). All assembly processes were carefully performed to avoid applying extra bending tension to the fibers.

\section{Bench Test of Assembled Spiral Fiber Tracker}

\begin{figure}[t] 
\centering 
\includegraphics[width=0.451\textwidth,keepaspectratio]{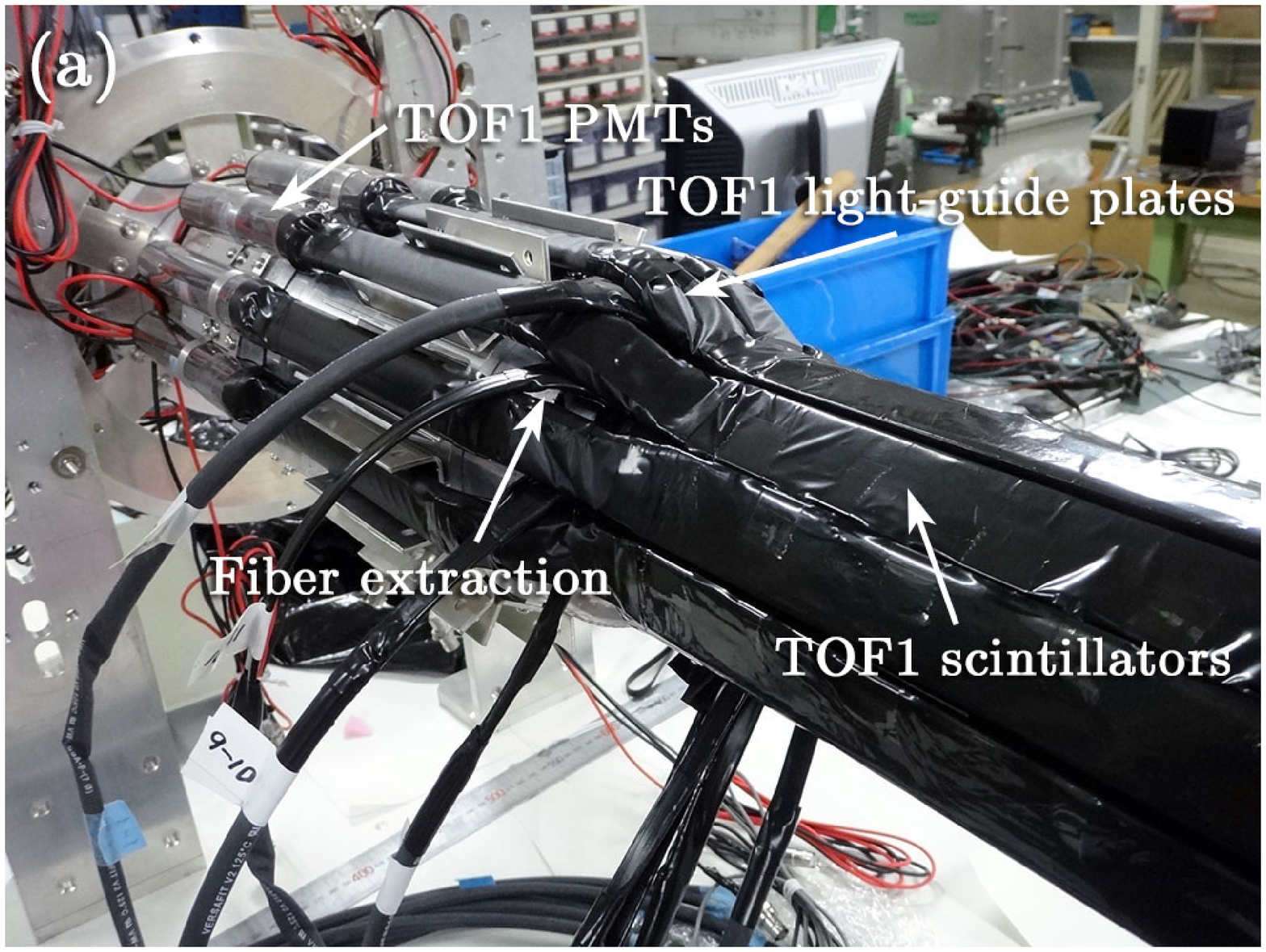}
\includegraphics[width=0.451\textwidth,keepaspectratio]{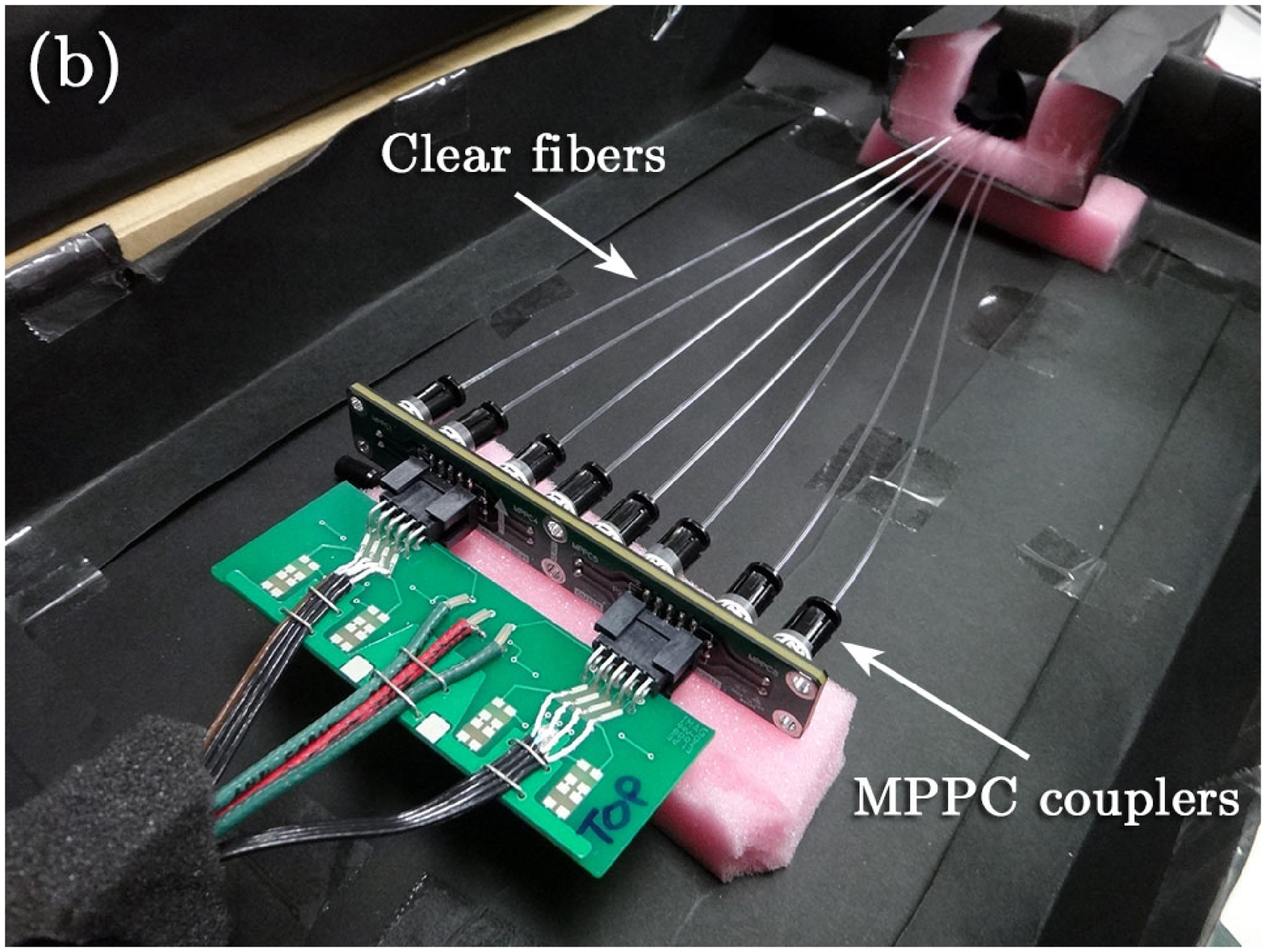}
\caption{(a) SFT and TOF1 counters attached to the central support structure. The SFT-fiber bundles were extracted from gaps between adjacent TOF1 light-guide plates. (b) Fiber readout experimental setup in the light-shielded box. The clear optical fibers extracted from the SFT body were connected to the coupler containing MPPCs.}
\label{fig:fig5}
\end{figure}

After installing the TOF1 counters around the ribbon body, in July 2014, we examined the SFT for dead fibers. Eleven TOF1 counters, except for the lowest (sixth) counter, were attached along TOF1 support structures while paying attention to the extraction of the SFT-fiber bundle from adjacent TOF1 counters (Fig. \ref{fig:fig5}(a)). A 2-mm-thick scintillation counter replaced the sixth TOF1 counter to trigger $\beta $ rays generated from another strontium-90 radioisotope positioned lower than the scintillation counter. Each clear fiber end glued to the positive MPPC coupler was connected to each negative MPPC coupler, which contains an MPPC. As shown in Fig. \ref{fig:fig5}(b), the MPPC module was set up in a light-shielded box at room temperature. The MPPCs were read out using a VME-compatible front-end board with the extended analogue SiPM integrated readout chip (EASIROC) \cite{cite13}. Because of these measurements, we confirmed that the MPPCs were detecting signals from all fibers, and therefore concluded that the SFT had no dead channels. After the sixth TOF1 counter was attached, the central detector module containing the SFT was transferred to J-PARC. We are now calibrating the entire SFT system using cosmic rays. In engineering runs prior to physics runs, the SFT will be calibrated more precisely for measuring the detection efficiency using MWPCs.

\section{Conclusions}

We plan to employ a new charged particle tracking detector in the TREK/E36 experiment scheduled at J-PARC. The device, called the SFT, consists of spirally coiled multilayer ribbons made of 1-mm-diameter plastic scintillating fibers and MPPCs; the SFT is designed to detect particle hits by reading out scintillation photons. An initial bench test of a 1.5-m-long prototype of two-layer fiber ribbon in the staggered configuration indicated that the tracker would basically have a detection efficiency exceeding 99\% for a three-photoelectron threshold using PMT readout (as an alternative to MPPC). The actual SFT was completely assembled around the kaon stopping target holder. In a post-assembly measurement, no dead channel was determined.

\section*{Acknowledgments}

We are grateful to the members of the TREK/E36 collaboration for their assistance. We are also grateful to Moderation-Line Co., Ltd., Japan, for their contributions to produce the fiber ribbon. This work was partly supported by the Russian Foundation Grant \#14-12-00560. The Canadian group was supported by the Natural Sciences and Engineering Research Council (NSERC) of Canada. M. Tabata was supported partly by the Space Plasma Laboratory at the Institute of Space and Astronautical Science (ISAS)/Japan Aerospace Exploration Agency (JAXA).

\end{document}